\documentclass[epsf,twocolumn,nofootinbib,notitlepage,superscriptaddress]{revtex4-1}

\usepackage[utf8]{inputenc}
\usepackage{bm}
\usepackage{graphicx}
\newcommand{\cnot}{{\textsc{CNOT}}}
\newcommand{\swap}{{\textsc{SWAP}}}
\newcommand{\rotation}{{\textsc{R}}}
\newcommand{\hadamard}{{\textsc{H}}}

\newcommand{\etal}{{\it et al.}~}
\usepackage{bbold}
\usepackage{color}
\usepackage{tikz}
\usepackage{float}
\usepackage{subcaption}
\usepackage{amsmath}
\usepackage{ulem}
\usepackage{xpatch}
\usepackage{tipa}
\makeatletter
\xpatchcmd{\@ssect@ltx}{\@xsect}{\protected@edef\@currentlabelname{#8}\@xsect}{}{}
\xpatchcmd{\@sect@ltx}{\@xsect}{\protected@edef\@currentlabelname{#8}\@xsect}{}{}
\makeatother
\makeatletter
\renewcommand*{\@fnsymbol}[1]{\ensuremath{\ifcase#1\or *\or \dagger\or \ddagger\or
   \mathsection\or \mathparagraph\or \|\or **\or \dagger\dagger
   \or \ddagger\ddagger \else\@ctrerr\fi}}
\makeatother

\usepackage{hyperref}
\usepackage{nameref}
\begin{document}

\title{Scaling Quantum Approximate Optimization on Near-term Hardware}

\author{Phillip C.~Lotshaw}
\email{lotshawpc@ornl.gov}
\thanks{\\ This manuscript has been authored by UT-Battelle, LLC under Contract No. DE-AC05-00OR22725 with the U.S. Department of Energy. The United States Government retains and the publisher, by accepting the article for publication, acknowledges that the United States Government retains a non-exclusive, paid-up, irrevocable, world-wide license to publish or reproduce the published form of this manuscript, or allow others to do so, for United States Government purposes. The Department of Energy will provide public access to these results of federally sponsored research in accordance with the DOE Public Access Plan. (http://energy.gov/downloads/doe-public-access-plan)}
\affiliation{
	Quantum Computational Sciences Group, Oak Ridge National Laboratory, Oak Ridge, Tennessee 37830 USA}

\author{Thien Nguyen}
\affiliation{
	Beyond Moore Computing Group, Oak Ridge National Laboratory, Oak Ridge, Tennessee 37830 USA}
\affiliation{
	Quantum Science Center, Oak Ridge National Laboratory, Oak Ridge, Tennessee 37830 USA}
\author{Anthony Santana}
\affiliation{
	Beyond Moore Computing Group, Oak Ridge National Laboratory, Oak Ridge, Tennessee 37830 USA}
\author{Alexander McCaskey}
\affiliation{
	Beyond Moore Computing Group, Oak Ridge National Laboratory, Oak Ridge, Tennessee 37830 USA}
\affiliation{
	Quantum Science Center, Oak Ridge National Laboratory, Oak Ridge, Tennessee 37830 USA}
\author{Rebekah Herrman}
\affiliation{
	Department of Industrial and Systems Engineering, University of Tennessee, Knoxville, Tennessee  37996-2315 USA}
\author{James Ostrowski}
\affiliation{
	Department of Industrial and Systems Engineering, University of Tennessee, Knoxville, Tennessee  37996-2315 USA}
\author{George Siopsis}
\affiliation{
	Department of Physics and Astronomy, University of Tennessee, Knoxville, Tennessee  37996-1200 USA}
\author{Travis S.~Humble}
\affiliation{
	Quantum Computational Sciences Group, Oak Ridge National Laboratory, Oak Ridge, Tennessee 37830 USA}
\affiliation{
	Quantum Science Center, Oak Ridge National Laboratory, Oak Ridge, Tennessee 37830 USA}
\date{\today}
\begin{abstract}
The quantum approximate optimization algorithm (QAOA) is an approach for near-term quantum computers to potentially demonstrate computational advantage in solving combinatorial optimization problems.  However, the viability of the QAOA depends on how its performance and resource requirements scale with problem size and complexity for realistic hardware implementations.  Here, we quantify scaling of the expected resource requirements by synthesizing optimized circuits for hardware architectures with varying levels of connectivity.  Assuming noisy gate operations, we estimate the number of measurements needed to sample the output of the idealized QAOA circuit with high probability.  We show the number of measurements, and hence total time to solution, grows exponentially in problem size and problem graph degree as well as depth of the QAOA ansatz, gate infidelities, and inverse hardware graph degree.  These problems may be alleviated by increasing hardware connectivity or by recently proposed modifications to the QAOA that achieve higher performance with fewer circuit layers.  \end{abstract}
\maketitle
\section*{Introduction}
Combinatorial optimization problems are commonly viewed as a potential application for near-term quantum computers to obtain a computational advantage over conventional methods \cite{Preskill2018quantum}.  A common approach to solving these problems uses the quantum approximate optimization algorithm (QAOA) \cite{farhi2014quantum}, which begins with a ``cost" Hamiltonian typically defined as
\begin{equation} \label{C} C = \sum_i h_i Z_i + \sum_{i,j} J_{i,j} Z_i Z_j \end{equation}
with real coefficients $J_{i,j}$ and $h_i$ that encode a quadratic unconstrained binary optimization problem in the eigenspectrum of $C$ \cite{Lucas2014qubo}. The QAOA prepares a quantum state $\vert \bm{\gamma}, \bm{\beta}\rangle$ on $n$ qubits using $p$ layers of unitary operators, where each layer alternates between Hamiltonian evolution under $C$ and under a ``mixing" Hamiltonian $B = \sum_{i=1}^n X_i$ composed of independent Pauli-X operators,
\begin{equation} \label{QAOA} \vert \bm{\gamma}, \bm{\beta}\rangle = \left(\prod_{l=1}^p e^{-i \beta_l B}e^{-i \gamma_l C}\right) \vert + \rangle^{\otimes n}. \end{equation}
The state is then measured to yield the $n$-bit binary string $z$ as a candidate solution to the problem.  The angles  $\bm{\beta} = (\beta_1,...,\beta_p)$ and $\bm{\gamma} = (\gamma_1,...,\gamma_p)$ are variational parameters chosen to minimize or maximize the expectation value $\langle C \rangle = \langle\bm{\gamma}, \bm{\beta} \vert C \vert \bm{\gamma}, \bm{\beta}\rangle$, depending on whether the optimal solution in $C$ is the minimum or maximum value, respectively.
\par
Farhi \etal have argued that QAOA recovers the ground state of $C$ as $p \to \infty$ \cite{farhi2014quantum}, but the primary interest in QAOA is in reaching high performance with a modest number of layers $p$ that could realistically be implemented on a quantum computer.  A significant body of theoretical 
\cite{Wurtz2021Bounds, wang2018quantum, Shaydulin2020Symmetries, Hadfield2018dissertation, Hadfield2021framework}, 
computational 
\cite{Galda2021transfer, zhou2020quantum, ReachabilityDeficit, shaydulin2019multistart, Shaydulin2020CaseStudy}, and experimental 
\cite{Google2021QAOA,Pagano25396} 
research has focused on understanding QAOA performance at $p\approx 1$, mostly on the MaxCut problem with a small number of qubits $n$, but also for other types of problems \cite{Pontus2020tail,Szegedy2020GraphQAOA,Harwood2021routing}.  
These studies have shown some promising results, for example, with QAOA outperforming the conventional lower bound of the GW algorithm for MaxCut on some small instances \cite{crooks2018performance, Lotshaw2021BFGS}. There have also been a variety of proposed modifications to the algorithm to improve performance \cite{herrman2021ma, Gupta2020WarmStart, zhu2020adaptqaoa, Egger2021warmstart, Wurtz2021spanningtree,farhi2017hardware,Patti2021nonlinearqaoa, LiLi2020Gibbs} and solve optimization problems with constraints \cite{hadfield2019quantum,Eidenbenz2020GroverMixers,Bartschi2020MaxkCover}.  The results from these and other studies have encouraged research into extending the QAOA to larger and more complex problems.
\par
In contrast to the QAOA studies focused on a small number of variables $n$, conventional computational methods are capable of handling problem instances with hundreds of variables or more.  To assess the usefulness of QAOA it will be necessary to scale to larger and more complex instances where it can be directly compared against these methods on practically relevant problems.  A recent study suggests that hundreds of qubits are needed \cite{guerreschi2019qaoa} to compete in time-to-solution, while the theoretical and experimental performance in this context are important open questions.  Theoretical considerations indicate that the number of layers $p$ will need to scale at least as $\log(n)$ in some instances, as the locality of the ansatz limits the ability to build global correlations that are needed for globally optimal solutions \cite{Farhi2020seegraph,Farhi2020seegraph2}.  Classical algorithms have also been developed that outperform QAOA at low $p$ \cite{hastings2019classical, marwaha2021local}, further suggesting large $p$ may be necessary to compete with conventional methods.  To optimize parameters at large $n$ and $p$, a variety of computational \cite{Lykov2020tensorqaoa,Medvidovic2021QAOA54qubit} and theoretical \cite{brandao2018concentration,wurtz2021cd,wurtz2021fixedangle,biamonte2021concentration,biamonte2021progress,Basso2021advantage} approaches have been developed and in some cases the theoretical performance has been characterized.  With parameter setting strategies at hand, what remains to be seen is how the QAOA will perform in experimental implementations.  The prospect of experimentally implementing the QAOA at large $n$ and $p$ raises questions about how quantum computing resources will scale with problem size and complexity, and how noise will influence the behavior of the algorithm.
\par
Here we report on the scaling of resources needed by QAOA on near-term intermediate-scale quantum (NISQ) devices. We show how features of the combinatorial problem and the target hardware influence the total number of gates and measurements required to reach a specified threshold of accuracy. First we consider problem features such as the average degree $d_G$ of the graph defining the problem instance, where $d_G$ is related to the number of non-zero terms in the quadratic unconstrained binary optimization problem. While much of the QAOA literature has focused on problems with small $d_G$, larger $d_G$ arises naturally in constrained combinatorial optimization problems \cite{herrman2021lower,Herrman2021GVS}.  In addition to $d_G$, the problem size $n$ and the number of QAOA layers $p$ also contribute to the gate counts and hence the resources required to implement the algorithm. It is furthermore important to consider the constraints that arise in current NISQ hardware due to limited connectivity on the hardware device qubit register, which can require costly $\swap$ gates to transport logical qubits. We show that the interplay between these logical requirements and hardware constraints generate steep  scaling in the resources required for high-fidelity implementation of QAOA as $n$, $p$, and $d_G$ increase.
\par
\begin{figure}
 \centering
    \begin{subfigure}{.2\textwidth}
     \includegraphics[width=3cm,height=7cm,keepaspectratio]{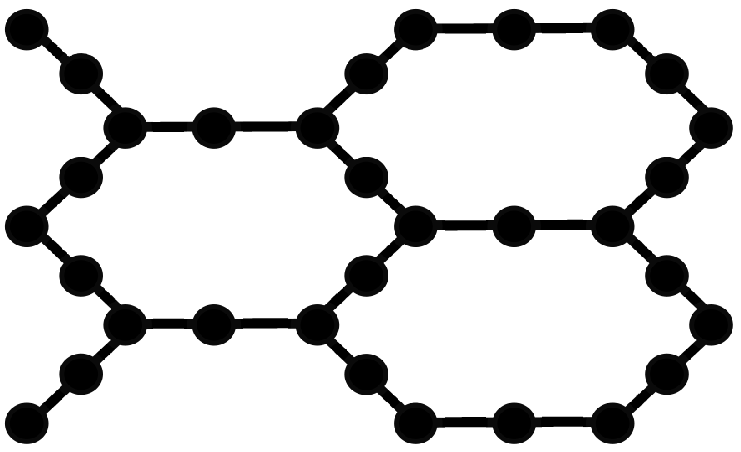}
     \caption{}
     \end{subfigure}
     \begin{subfigure}{.2\textwidth}
     \includegraphics[width=3cm,height=7cm,keepaspectratio]{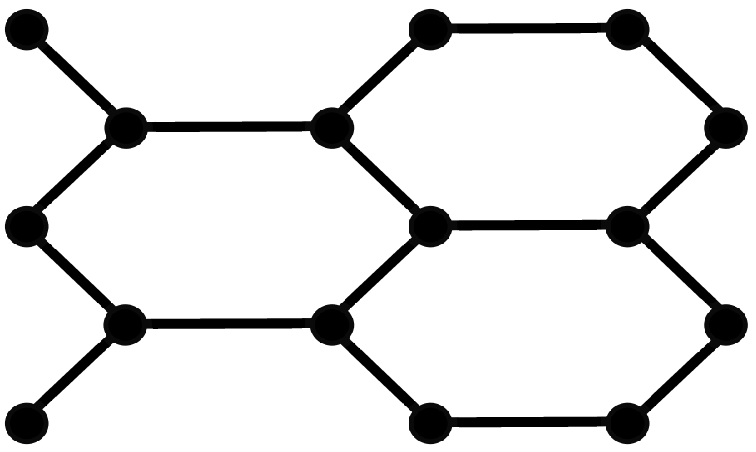}
     \caption{}
     \end{subfigure}
     \begin{subfigure}{.2\textwidth}
     \includegraphics[width=2.cm,height=7cm,keepaspectratio]{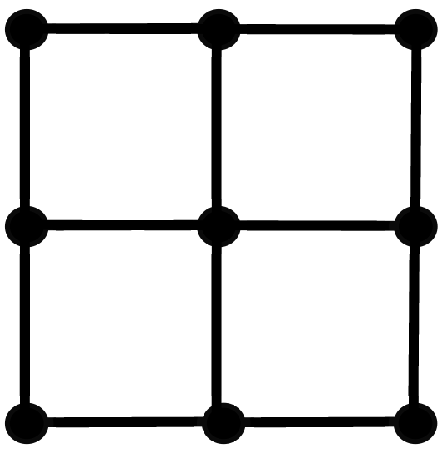}
     \caption{}
     \end{subfigure}
     \begin{subfigure}{.2\textwidth}
     \includegraphics[width=3cm,height=7cm,keepaspectratio]{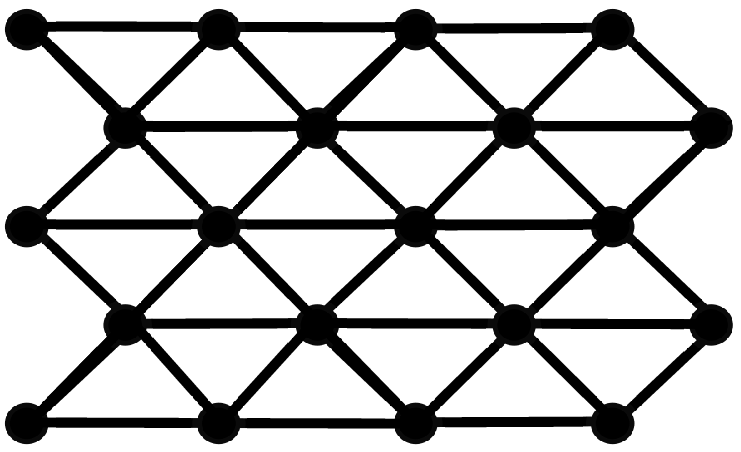}
    \caption{}
     \end{subfigure}
     \caption{Hardware connectivity graphs for (a) heavy-hexagon, $d_H=2.5$ (b) hexagon, $d_H=3$, (c) square, $d_H=4$, and (d) triangle, $d_H=6$.}
     \label{hardware}
\end{figure}

 
Our approach synthesizes optimized circuit representations of QAOA for varying problem sets targeting constrained noisy hardware. We optimize both the number of gates and the overall performance through judicious placement of the logical qubits and injected $\swap$ gates. Placement and routing are difficult optimization problems and it is not clear {\it a priori} how an ideal QAOA instance expressed as Eq.~(\ref{QAOA}) will map to a given hardware \cite{Li2019tackling,Siraichi2019Enfield,Zulehner2018mapping,Nannicini2021assignment}.  To understand the role of hardware connectivity, we synthesized optimized QAOA circuits on scaled versions of each of the connectivity architectures shown in Fig.~\ref{hardware}. These planar architectures correspond to contemporary and hypothesized hardware designs.  Each architecture has a distinct connectivity defined as the average hardware graph degree $d_H$, i.e., the average number of distinct two-qubit gate connections per hardware register element (ignoring perimeter elements to give a size-independent $d_H$). The architectures range from $d_H = 2.5$ for the heavy hexagonal lattice in Fig.~\ref{hardware}(a) to $d_H = 6$ for the triangular lattice in Fig.~\ref{hardware}(d). We quantify the $\swap$ gate counts with respect to $d_H, d_G, n,$ and $p$, and we fit scaling relations to these results.  
\par
Resource counts also give insight into the scalability of the QAOA in the presence of noise.  We define a simple noise model for a quantum state traversing a circuit with gate counts estimated from our resource analysis and use this to quantify the reliability of QAOA as it scales to larger and more complex problems.  Our analysis complements previous theoretical results describing how noise influences the QAOA cost expectation value, trainability, and eigenvectors of the density operator \cite{Coles2020nibp, Quiroz2021precision, Xue2021noise, Koczor2021dominant}. We quantify the number of measurements $M$ that are needed to obtain a single result from the idealized state that would be produced by a noiseless version of the circuit.  This characterizes the reliability of the algorithm and the expected time-to-solution $T$, assuming $T \propto M$.  The results assess the scalability of the QAOA on noisy near-term hardware and the expected influence of $d_H, d_G, n,$ and $p$.
\section*{Results}
\subsection*{Mapping to Hardware}
We express the QAOA unitary operators of Eq.~(\ref{QAOA}) in terms of a hardware gate set of Hadamards $\hadamard$, $Z$-rotations $\rotation(\theta)$, and controlled-\textsc{NOT} $\cnot$, as described in \nameref{methods}. The gate-to-unitary operator correspondences given there provide the minimal numbers of each type of gate that must be implemented in the algorithm, for example, on fully connected hardware.  
\par
It is useful to classify problem instances $C$ in terms of their circuit structure. We define problem graphs $G$ with vertices for each qubit $i$ and edges $\langle i,j\rangle$ for each non-zero $J_{i,j}$ constant in Eq.~(\ref{C}).  Each edge $\langle i,j\rangle$ requires a set of two-qubit gates $\cnot_{i,j} \rotation_j(2J_{i,j}\gamma_l)\cnot_{i,j}$ and the total set of edges defines all two-qubit gates that are needed on fully connected hardware.  The specific values of the parameters $J_{i,j} \neq 0$, $h_i$, $\gamma_l$, and $\beta_l$ enter as rotation angles in the circuit, hence all problem instances with the same problem graph have the same circuits up to choices of these angles.  When an $h_i=0$ then a single-qubit gate can be further removed from the circuit, but this does not affect the two-qubit gate structure. We consider all non-isomorphic connected problem graphs with $n=7$ qubits to determine how the circuits scale with the average problem graph degree $d_G$; to determine scaling with the number of qubits we assess 3-regular problem graphs with $d_G=3$ at varying $n$.   On fully connected hardware, the number of gates of each type are
\begin{align} 
\label{NH} N_\hadamard &= 2np+n,\\
\label{NR} N_\rotation &= p\left(\eta + \frac{n(d_G+2)}{2} \right),\\
\label{NCNOT connected} N_\cnot^\mathrm{fc} &= pnd_G - N_0,
\end{align}
where $\eta$ is the number of non-zero $h_i$ in Eq.~(\ref{C}) and $N_0 \leq \lfloor n/2\rfloor$ is an instance-dependent number of $\cnot$ gates that can be removed from the first layer of the circuit as they do not affect the initial state \cite{Majumdar2021MaxcutOpt}, see Supplemental Information Sec.~I for details.
\par
However, on hardware with limited connectivity, it is often the case that some of the two-qubit gates cannot be implemented by any initial placement of the logical qubits onto the hardware register.  For example, a non-planar problem graph cannot be mapped onto any of the planar registers in Fig. \ref{hardware}.  It is therefore necessary to use $\swap$ gates to shuttle logical qubits around the register during execution of the circuit, to realize connections that are not available to the initial qubit placement.  There are many potential circuits that can be created and these can result in different total numbers of $\swap$ gates, with up to ${n \choose 2}$  $\swap$ gates in $n$ circuit layers in the worst case \cite{OGorman2019swapnetwork,Kivlichan2018lineardepth}.  An ideal circuit will minimize the number of gates or circuit depth to minimize the negative impacts of noise in the circuit.
\par
We compute circuits that minimize $\cnot$ gate counts for each register architecture in Fig. \ref{hardware} using an optimization routine. We optimize single layers of the QAOA algorithm as additional layers have the same circuit structure apart from differences in the qubit locations due to $\swap$ gates.  These differences can be accounted for by mirroring the circuit implementation of $\exp(-i\gamma_l C)$ in subsequent layers, so that qubits move back and forth between locations from layer to layer.  For an $n$-qubit problem instance, we use register grids of sizes just larger than $\sqrt{n} \times \sqrt{n}$, as we found that further increasing the grid size tended to result in larger optimized circuits.  Our optimization procedure uses two nested loops.   The inner loop calls the circuit mapping algorithm SABRE \cite{Li2019tackling}, which generates a set of random placements of the logical qubits onto the hardware register then optimizes each placement, ultimately returning the final optimized circuit with the smallest depth.  For our circuits, we have found that SABRE sometimes yields sub-optimal placements, as it does not recognize the commutativity of the terms $\exp(-i \gamma_l J_{i,j} Z_i Z_j)$ in Eq.~(\ref{QAOA}), but instead tries to implement these in the order it is given.  We therefore define an outer loop that randomly shuffles these commuting terms, to optimize over varying term orderings.  This outer loop decreases the number of gates in our optimized circuits compared to a more basic implementation with SABRE only. For each problem graph, we take our final result from these nested loops as the circuit with the fewest $\cnot$ gates.  The total number of $\cnot$ gates on hardware with limited connectivity with $N_\swap$ $\swap$ gates is
\begin{equation}\label{NCNOT}  N_\cnot = N_\cnot^\mathrm{fc}+ p\sigma N_\swap,\end{equation}
where $\sigma$ quantifies the average increase in $\cnot$ gates per $\swap$ gate, beyond the $N_\cnot^\mathrm{fc}$ gates that are needed on fully connected hardware.  Each $\swap$ gate is defined as a product of three $\cnot$ gates, so $\sigma=3$ in the worst case.  In better cases, a $\swap_{ij}$ gate is placed adjacent to a $\cnot_{ij}$ gate in the circuit and $\cnot_{ij}\cnot_{ij}=\mathbb{1}$ is used to remove a pair of gates.  This gives $1 \leq \sigma \leq 3$ in our accounting.  Further details of the implementation, convergence behavior, and performance can be found in Supplemental Information Sec.~I.
\subsection*{Scaling with Problem Size and Degree}
\begin{figure}
    \centering
        \includegraphics[width=3.5cm,height=9cm,keepaspectratio,trim={4cm 0cm 4cm 0.25cm }]{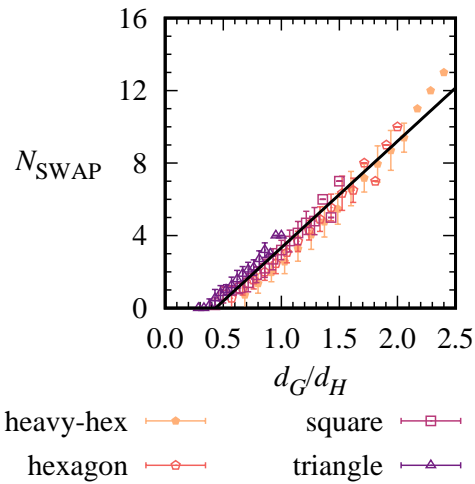}
    \caption{$\swap$ gate scaling with average problem degree $d_G$ and hardware degree $d_H$ for 7-vertex graphs.  The solid line shows the non-linear least squares fit to $N_\swap(d_G,d_H) = a d_G/d_H + b$, with $a=5.9 \pm 0.1$ and $b = -2.5 \pm 0.2$, with $\pm$ indicating the asymptotic standard error of the fit parameters.}
    \label{n=7 results}
\end{figure}

We next mapped circuits for each of the 853 non-isomorphic problem graphs at $n=7$ \cite{GraphFiles}. The results in Fig.~\ref{n=7 results} show how the number of $\swap$ gates $N_\swap$ scales with the average problem graph degree $d_G$ at this $n$ across our hardwares with varying $d_H$.  As $d_G$ increases so does the number of edges in the graph, and hence the number of two-qubit gates in each layer of the QAOA algorithm.  Greater numbers of $\swap$ gates are needed on average to accommodate these two-qubit gates.  Similarly, as the hardware degree $d_H$ increases a greater number of two-qubit gates are available natively on the hardware, so fewer $\swap$ gates are needed.  The mean numbers of $\swap$ gates at each $d_G$ and $d_H$ are fit by an empirical linear relation $N_\swap(d_G,d_H) \sim d_G /d_H$ with fit parameters in the figure caption and a root-mean-square-error (RMSE) of 0.58 $\swap$ gates.  The small error indicates the empirical relation is successful in providing a unified account of the $N_\swap$ scaling across problem graphs and hardware architectures at this $n$.
\par
Next we consider how the number of $\swap$ gates scales with the size of the problem $n$.  We considered sets of 3-regular graphs with 108 graph instances each at $n=20,40,$ and $60$ qubits.  The $3$-regular problem graphs have three non-zero $J_{i,j}$ terms for each qubit $i$ in Eq.~(\ref{C}) and this standardizes $d_G=3$ as we scale to larger sizes.  Three-regular graphs have also been studied with considerable interest in the QAOA MaxCut literature \cite{farhi2014quantum, guerreschi2019qaoa, Wurtz2021Bounds, wurtz2021fixedangle, brandao2018concentration, zhou2020quantum, Galda2021transfer} and in a previous experimental demonstration of QAOA \cite{Google2021QAOA}.  They are appealing targets for near-term hardware since most graphs at the same $n$ have higher average degree $d_G$, hence we expect them to require more noisy two-qubit gates, due to both the increase in the minimal number of $\cnot$ gates in Eq.~(\ref{NCNOT connected}) and also the expected increase in $\swap$ gates following the previous analysis of Fig.~\ref{n=7 results}.
\par
\begin{figure}
     \includegraphics[width=7.5cm,height=9cm,keepaspectratio,trim={1cm 0 1cm 0cm },clip]{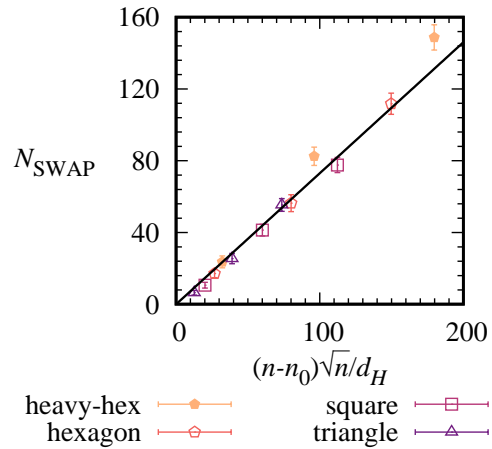}
     \caption{Average $\swap$ gate scaling with number of qubits $n$ and hardware degree $d_H$ for 3-regular graphs.}
     \label{3 regular mapping}
\end{figure}

We computed optimized circuit mappings for these 3-regular instances to obtain the key result pictured in Fig.~\ref{3 regular mapping}, which relates the number of $\swap$ gates to the average hardware degree $d_H$ as the problem size $n$ increases.   We fit the data with an empirical curve that is based on counting the number of two-qubit terms that cannot be implemented by the initial qubit placement and assuming the number of $\swap$ gates needed to bring the qubits together for  these edge terms increases on average in proportion to the length and width of the hardware grid, see \nameref{methods} for details. This leads to the empirical relation shown by the solid line in the figure
\begin{equation} \label{Nswap 3 regular} N_\swap(n,d_H) = \mu (n-n_0)\sqrt{n} / d_H , \end{equation}
where $\mu=0.73\pm 0.02$ is a fit parameter computed through non-linear least squares and $\pm 0.02$ is the asymptotic standard error.  Here $n_0$ sets the zero of $N_\swap$ and represents the maximum problem graph size at which all graphs can be mapped to hardware, for example, for fully connected hardware $n_0=n$ and $N_\swap(n,d_H)=0$.  For the triangle lattice in Fig. \ref{hardware}(d), all 3-vertex problem graphs can be mapped directly onto the lattice but the 4-vertex complete graph cannot be, so $n_0=3$. For the other hardware lattices, $n_0=2$.  
\par
We assess the performance of the empirical formula using the RMSE between the average $N_\swap$ and the empirical $N_\swap(n,d_H)$.  Across all results in Fig.~\ref{3 regular mapping}, the RMSE=7.2 $\swap$ gates. The RMSE is strongly influenced by the outliers for the heavy-hexagon array at $n=40$ and $n=60$, where the empirical formula is up to 16\% smaller than the results.  These deviations may be related to the bimodal degree structure of the heavy-hexagon array in Fig.~\ref{hardware}(a), which has a mixture of register elements of degrees two and three, unlike the other constant-degree hardwares. Excluding the results for the heavy-hexagon at $n=40$ and $n=60$ decreases the RMSE to 2.7 $\swap$ gates.  We conclude the empirical formula is giving a good fit to the majority of data in the figure, apart from the heavy-hexagon at large $n$, where the formula gives a looser bound to the observed $N_\swap$. 
\subsection*{Noisy Architecture Model and Measurement Count Scaling}
We use a simple noise model for our circuits to assess how noise influences the scalability of the QAOA, in terms of the number of measurements $M$ that are needed from a noisy circuit to obtain a single result from the intended noiseless quantum state distribution.  This quantifies the reliability of a noisy QAOA circuit in producing the intended output and also characterizes the scaling in the time-to-solution $T$ assuming $T \propto M$. 
\par
An instance of a QAOA circuit is expressed in terms of a series of gates with ideal unitary evolution operators $U_0, U_1, \ldots$, with $U_\alpha \in \{\hadamard,\rotation,\cnot\}$ the unitary for the $\alpha$th gate, acting on an initial state $\rho_0 = (\vert 0\rangle\langle 0\vert)^{\otimes n}$.  The noisy state produced by the $\alpha$th gate is expressed using a quantum channel as
\begin{equation}\label{channel}
    \rho_{\alpha+1} = (1-\epsilon_\alpha)U_\alpha \rho_\alpha U_\alpha^\dag + \sum_{k=1}^K \epsilon_\alpha^{(k)} E^{(k)}_\alpha U_\alpha \rho_{\alpha} U_\alpha^\dag {E^{(k)}_\alpha}^\dag,
\end{equation}
where the Kraus operators $(\epsilon_\alpha^{(k)})^{1/2}E^{(k)}_\alpha$ give noisy deviations from the intended evolution with probabilities $\epsilon_\alpha^{(k)}$.  The final state of the circuit is \cite{Koczor2021dominant}
\begin{equation}\label{rho}
\rho = F_0 \rho_\mathrm{ideal} + (1-F_0)\rho_\mathrm{noise}
\end{equation}
where $\rho_\mathrm{ideal} = \vert \bm \gamma, \bm \beta \rangle\langle \bm \gamma, \bm \beta \vert$ is the density operator for the intended pure state $\vert \bm \gamma, \bm \beta \rangle$, $\rho_\mathrm{noise}$ is a density operator composed of all terms with at least one Kraus operator, and $F_0 = \prod_\alpha (1-\epsilon_\alpha)$ is a lower bound to the state preparation fidelity $F = \langle \bm \gamma, \bm \beta \vert \rho \vert \bm \gamma, \bm \beta \rangle \geq F_0$, with equality when $\mathrm{Tr} \rho_\mathrm{ideal}\rho_\mathrm{noise}=0$. If we assume constant error rates $\epsilon_\cnot$, $\epsilon_\hadamard$, and $\epsilon_\rotation$ for each $\cnot$, $\hadamard$, and $\rotation$ gate respectively, then
\begin{equation}\label{F0}
F_0 = (1-\epsilon_\cnot)^{N_\cnot}(1-\epsilon_\hadamard)^{N_\hadamard}(1-\epsilon_\rotation)^{N_\rotation}, 
\end{equation}
where the $N$ are the corresponding gate counts. 
\par
A noisy implementation of QAOA will be effective when it can produce measurement results from the intended state distribution $\rho_\mathrm{ideal}$. In the absence of readout errors, a measurement projects the total state $\rho$ onto a computational basis state $\vert z \rangle$ that is the result of the measurement, with probability $P(z) = \langle z \vert \rho \vert z \rangle = F_0 P_\mathrm{ideal}(z) + (1-F_0)P_\mathrm{noise}(z)$.  This has a lower bound $P(z) \geq F_0 P_\mathrm{ideal}(z)$ independent of the specific noise process, apart from the values of the error rates $\epsilon_\alpha$ that determine $F_0$.  Summed over all $\vert z \rangle$ in the support $S$ of $\rho_\mathrm{ideal}$, the total probability $P = \sum_{\vert z\rangle \in S} P(z)$ to obtain any result from the ideal state distribution is 
\begin{equation} 
P \geq F_0. 
\end{equation}
We use this probability inequality to bound the number of measurements $M = \log(1-\mathcal{P})/\log(1-P)$ that are needed to obtain a single sample from the distribution of the intended state with probability $\mathcal{P}$ \cite{Lotshaw2021BFGS,Pontus2020tail}, 
\begin{equation} 
M \leq \frac{\log(1-\mathcal{P})}{\log(1-F_0)}. 
\end{equation}
\begin{figure}
    \includegraphics[width=9cm,height=11cm,keepaspectratio,trim={1cm 0 1.5cm 0cm },clip]{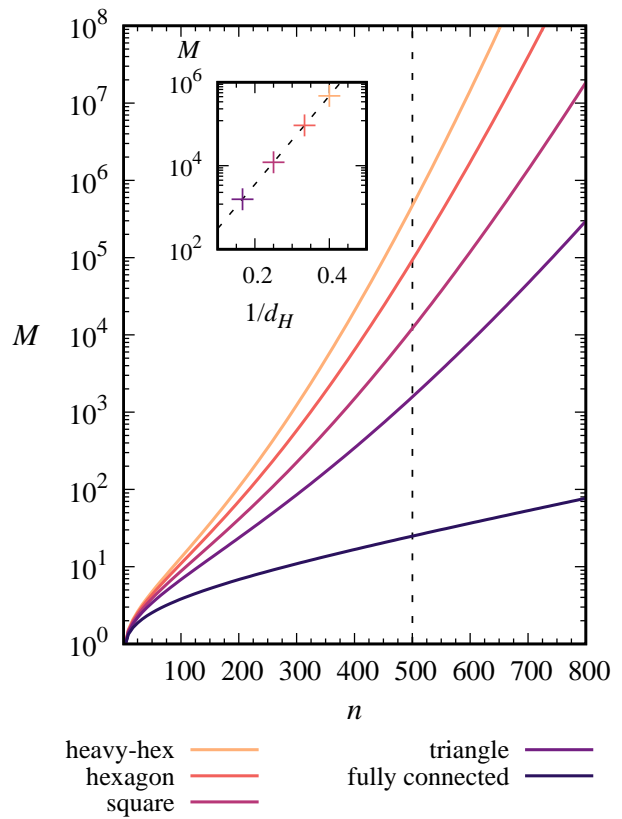}
     \caption{The number of measurement samples $M$ to measure a result from the intended state for 3-regular graphs, see text for details. Inset: $M$ diverges exponentially in $1/d_H$. }
     \label{M fig}
\end{figure}
\par
It is useful to consider a few examples.  In a theoretical best case of QAOA, the intended state is a single computational basis state $\vert \bm{\gamma}, \bm{\beta}\rangle = \vert z_\mathrm{opt}\rangle$ that gives the optimal cost value $C(z_\mathrm{opt}) = C_\mathrm{opt} \in \mathbb{R}$.  If we assume that noise does not contribute significantly to the probability for $\vert z_\mathrm{opt}\rangle$, then $P \approx F_0$ and $M$ is close to the upper bound.  In more generic cases of interest, the intended state has non-zero probability for a variety of approximately optimal states and the goal is to measure any one of these states.  In this case $M$ may be smaller than the upper bound, and potentially much smaller if the probability to measure approximately optimal states is significant for the $\rho_\mathrm{noise}$ component.  Smaller upper bounds for $M$ might then be obtained using information about the noise process and its expected influence in $\rho_\mathrm{noise}$. However, without detailed information about a specific state and noise process we do not have a way to decrease $M$ below the upper bound, which serves as a generic guide for any possible intended QAOA state and noisy evolution of the type in Eqs.~(\ref{channel})-(\ref{rho}).
\par
We assessed the scalability of the number of measurement samples by computing the upper bound for $M$ for 3-regular graphs at varying sizes $n$ and at $p=20$ QAOA layers, with a probability $\mathcal{P}=0.99$ to sample from the intended state distribution.  We consider 3-regular problem graph instances with gate counts $N_\hadamard$, $N_\rotation$, and $N_\cnot$ in Eqs.~(\ref{NH}),(\ref{NR}), and (\ref{NCNOT}) respectively, assuming all $h_i \neq 0$ in Eq.~(\ref{C}) so that $\eta=n$.  We use $N_\swap$ computed from the empirical formula of Eq.~(\ref{Nswap 3 regular}) for each hardware architecture in Fig.~\ref{hardware}, $\sigma=3$ as the number of additional $\cnot$ gates per $\swap$ gate in Eq.~(\ref{NCNOT}), in accord with our results at large $n$ from Supplemental Information Sec.~I, and we approximate $N_0 = 0$ since $N_0 \ll N_\cnot$ when $p=20$. The $F_0$ in $M$ is then computed from Eq.~(\ref{F0}) with assumed error rates of $\epsilon_\cnot = 5 \times 10^{-5}$ and $\epsilon_\rotation = \epsilon_\hadamard = \epsilon_\cnot/10$. For comparison, recent advances in transmon qubits have achieved two-qubit gate error rates of $6.4 \times 10^{-3}$ and single-qubit error rates of $3.8 \times 10^{-4}$ \cite{Jurcevic2021QV64}. 
\par
Figure \ref{M fig} shows how this $M$ scales with problem size $n$.  The number of measurements increases exponentially with $n$ at a rate that depends on the hardware degree $d_H$.  The variations in hardware themselves give an exponential divergence in $M$ as the reciprocal hardware degree $1/d_H$ increases and the hardware becomes less connected (Fig.~\ref{M fig} inset), due to the empirical dependence of $N_\swap \sim 1/d_H$ from Eq.~(\ref{Nswap 3 regular}).  The hardware dependence is significant at the large $n$ that are required for practical problems. For example, at $n=500$ (vertical dotted line), the number of measurement samples is approximately $20$ for fully connected hardware but increases by four orders of magnitude going to the least connected hardware (heavy-hexagon, Fig.~\ref{hardware}(a)).  Here $n=500$ exemplifies a nontrivial problem size but is otherwise arbitrary---similar scaling behavior is observed for other large $n$. Curves similar to Fig.~\ref{M fig} can also be computed for fixed $n$ as the error rates $\epsilon_\alpha$, number of QAOA layers $p$, or as the problem graph degree $d_G$ increase, see Supplemental Information Sec.~III for details.
\section*{Discussion}
Prospects for obtaining a quantum computational advantage with the QAOA are expected to require hundreds of qubits or more to compete against conventional methods on practically relevant problems \cite{guerreschi2019qaoa, Harwood2021routing}. As the QAOA scales to larger and more complex problems, the number of gates to implement the algorithm on fully connected hardware increases with the problem graph degree $d_G$ and number of qubits $n$. For sparsely connected hardware additional $\swap$ gates are needed.  We computed optimized circuits to determine how the number of $\swap$ gates $N_\swap$ scales with $n$ and $d_G$ on a variety of real and hypothetical hardware architectures with varying levels of connectivity in terms of the hardware degree $d_H$.  The reciprocal hardware degree $1/d_H$, average problem graph degree $d_G$, and number of qubits $n$ were each found to be important scaling factors in the empirical behavior of $N_\swap$. Using a simple noise model with gate counts extrapolated from our circuits we computed the number of measurement samples $M$ from a noisy circuit that are needed to obtain a single measurement from the distribution of an idealized noiseless version of the state with probability $\mathcal{P}$.   This is a measure of the reliability of a noisy circuit in producing the intended outcome. We argued that $M$ increases exponentially with $n$, $d_G$, $1/d_H$, the number of QAOA layers $p$, and the gate error rates $\epsilon_\alpha$.  Assuming that $M$ is proportional to the time to solution, this corresponds to an exponential time complexity in each of these factors. 
\par
We considered $n=500$ as an example of a nontrivial problem size to compare the number of measurements across different hardwares.  Our results show that the number of measurement samples is $2 \times 10^3 \leq M \leq 5\times10^5$ at this $n$ and $p=20$ for the considered error rates and hardwares.  These numbers of measurements should not be difficult to obtain from a quantum computer.  However, our parameter choices and problem sets were optimistic in some respects.  The assumed error rates were about two orders of magnitude below current state of the art devices \cite{Jurcevic2021QV64} and larger error rates exponentially increase the number of measurements.  For example, doubling the error rates so that $\epsilon_\cnot = 10^{-4}$ gives $5 \times 10^{5} \leq  M \leq 5 \times 10^{10}$ for our hardwares.  We also assumed 3-regular problem graphs, which have been studied with great interest in the QAOA literature.  However, many practically relevant problems use denser problem graphs, for example in constrained optimization problems \cite{Herrman2021GVS,herrman2021lower,Harwood2021routing}.  For denser graphs the average degree can scale as $n$  and changes in degree can significantly affect $M$. For example, using our approach and parameter choices for a 500 qubit problem graph with average degree $d_G=25$ we obtain $M = 3\times10^{6}$ on fully connected hardware. For the sparsely connected hardware we consider we do not have a precise scaling relation for $N_\swap$ on $d_G=25$ graphs, but if we optimistically use the same relationship $N_\swap(n,d_H)$ we found for 3-regular graphs we obtain $2 \times 10^8 \leq M \leq 5 \times 10^{10}$ at $d_G=25$.  This is ignoring any dependence of $N_\swap$ on $d_G$, which would be significant if our small $n$ observation $N_\swap \sim d_G$ holds also at large $n$. A final note is that if more than one measurement is needed from the state with high probability, then this will introduce an additional scaling beyond the $M$ presented here. The numbers of measurements quickly become greater than what can realistically be expected from near-term quantum computers.

We expect the measurement scaling will significantly inhibit the ability to implement the QAOA at scales relevant for quantum advantage.  When the QAOA parameters are optimized using measurements from a quantum computer, this optimization will also be greatly inhibited. Parameter optimization has been addressed in some instances using theoretical approaches \cite{Galda2021transfer, crooks2018performance, Lotshaw2021BFGS, Lykov2020tensorqaoa, Medvidovic2021QAOA54qubit, brandao2018concentration, wurtz2021cd, wurtz2021fixedangle, biamonte2021concentration, biamonte2021progress, Basso2021advantage}, though for generic instances it is unclear if such approaches can be applied.  However, even with a good set of parameters the circuit must still be run to obtain the final bitstring solution to the problem, and in our model this requires a number of measurements that quickly becomes prohibitive at scales relevant for quantum advantage.  Straightforward attempts to scale the QAOA will face a significant barrier if these scaling problems are not addressed.
\par
Our expectations for performance are based on a general upper bound that is saturated when the noisy and ideal components of the total circuit density operator give distinct measurement results in the computational basis.  A vanishing overlap in measurement results is expected when the ideal QAOA circuit prepares a computational basis state, while intermediate superposition states may have non-negligible overlap with the noisy subspace.  Further analysis will require details from hardware-specific noise models to determine more precise estimates for how such errors influence $M$.  In addition, there are methods to overcome the measurement count limitations.  One approach is to significantly increase hardware connectivity or modify the gate set, for example, using ion-trap quantum computers with globally-entangling M{\o}lmer-S{\o}rensen gates \cite{Rajakumar2020unionofstars} or Rydberg atoms that naturally enforce constraints in some instances of QAOA \cite{Pichler2018RydbergQAOA}. Another approach is to modify the QAOA ansatz.  This includes introducing additional parameters within layers of QAOA \cite{herrman2021ma}, modifying the structure of the ansatz \cite{Wurtz2021spanningtree,Gupta2020WarmStart, Egger2021warmstart, zhu2020adaptqaoa}, modifying the cost function \cite{Patti2021nonlinearqaoa}, objective function \cite{LiLi2020Gibbs}, and circuit structure \cite{farhi2017hardware}.  Such technological and algorithmic advances are likely necessary to reduce the numbers of layers or gates, and hence the accumulated noise, as the QAOA scales to larger sizes.
 
\section*{Methods}\label{methods}
We generated circuits using the XACC quantum programming framework \cite{McCaskey2018XACC,McCaskey2019XACC} to map the unitary quantum operators of Eq.~(\ref{QAOA}) to a gate set of Hadamards $\hadamard$, $Z$-rotations $\rotation(\theta)=\exp(-i(\theta/2)Z)$, and controlled-NOT $\cnot$ gates. To map these circuits to hardware with limited connectivity, we used the Enfield software library \cite{Siraichi2019Enfield} and SABRE algorithm \cite{Li2019tackling} implemented within XACC.  Details of the implementation, convergence behavior, and comparison with a lower bound for $N_\swap$ at small $n$ are described in the Supplemental Information Sec.~I.
\par
In terms of our gate set, the unitary operators in Eq.~(\ref{QAOA}) are
\begin{align} 
\label{UC2}\exp(-i \gamma_l J_{i,j}Z_iZ_j) & = \cnot_{ij} \rotation_j(2\gamma_lJ_{i,j}) \cnot_{ij},\\
\exp(-i\gamma_lh_iZ_i) & = \rotation_i(2\gamma_lh_i)\\ 
\exp(-i\beta_l X_i) & = \hadamard_i \rotation_i(2\beta_l) \hadamard_i.
\end{align}
%
\subsection*{Empirical Formula for 3-regular Graphs}
We construct the empirical curve $N_\swap(n,d_H)$ in Eq.~(\ref{Nswap 3 regular}) by considering how many two-qubit gates cannot be implemented by the initial mapping of qubits onto the register along with the average expected behavior for how many $\swap$ gates are needed to bring qubits together for each of these gates.  We begin by separating the edge terms in a mapped problem graph instance into edges $s=\langle s_1, s_2\rangle$ that are ``satisfied" by the initial placement of qubits on the register, in the sense that the two-qubit gates between $s_1$ and $s_2$ can be implemented in the initial placement, and edges $u = \langle u_1, u_2 \rangle$ that are ``unsatisfied," in the sense that $\swap$ gates are needed to bring the qubits $u_1$ and $u_2$ together to implement their two-qubit gates. Our approach is to express the total number of $\swap$ gates as $N_\swap = \sum_u N_\swap^{(u)}$, where $N_\swap^{(u)}$ is the number of $\swap$ gates that are used in the circuit to bring qubits $u_1$ and $u_2$ together to implement the two-qubit gates for $u$.  
\par
Some care is needed to define the $N_\swap^{(u)}$ to give a consistent total $N_\swap$. Each $\swap$ gate moves locations of two qubits and hence can contribute to two terms $N_\swap^{(u)}$ and $N_\swap^{(u')}$;  one approach is to allow for fractional values in the $N_\swap^{(u)}$, for example, values of 1/2 in  $N_\swap^{(u)}$ and $N_\swap^{(u')}$ when a $\swap$ gate moves two qubits that help to satisfy $u$ and $u'$.  Another consideration is that a series of $\swap$ gates may be implemented before the gates for a given $u$, while along the way the $\swap$ gates that are relevant for $u$ may also allow for implementations of two-qubit gates for a variety of other $u',u'',...$.  We could then assign fractional values to each of the $N_\swap^{(u)}, N_\swap^{(u')}, N_\swap^{(u'')},...$ based on which qubits are moved by the series of $\swap$ gates and which unsatisfied edges they contribute to, such that $N_\swap = \sum_u N_\swap^{(u)}$.  A final consideration is that sometimes the circuits will $\swap$ qubits that are in initially satisfied edges $s$ before the two-qubit gates for those edges are implemented.  Although additional $\swap$ gates are sometimes used in these cases for the satisfied edges $s$, these $\swap$ gates are only needed because there were initially unsatisfied edges $u$ which began a series of $\swap$ gates earlier in the circuit, so it is reasonable to systematically assign the $\swap$ gates for these $s$ to the $N_\swap^{(u)}$.  Although the calculation of the $N_\swap^{(u)}$ is somewhat complicated by these considerations,  by design the total must always sum to $N_\swap$.  This can be expressed as an average $N_\swap = N_u \overline{N_\swap^{(u)}}$, where $N_u$ is the total number of unsatisfied edges and $ \overline{N_\swap^{(u)}}$ is the average number of $\swap$ gates per unsatisfied edge.  The $N_u$ is determined solely by the initial placement of qubits onto the register, while the average $\overline{N_\swap^{(u)}} = N_\swap/N_u$.  We argue for the behavior of these terms in determining $N_\swap$ and the empirical fit curve of Eq.~(\ref{Nswap 3 regular}).
\begin{figure}
    \includegraphics[width=8cm,height=11cm,keepaspectratio,trim={1cm 0cm 1cm 0},clip]{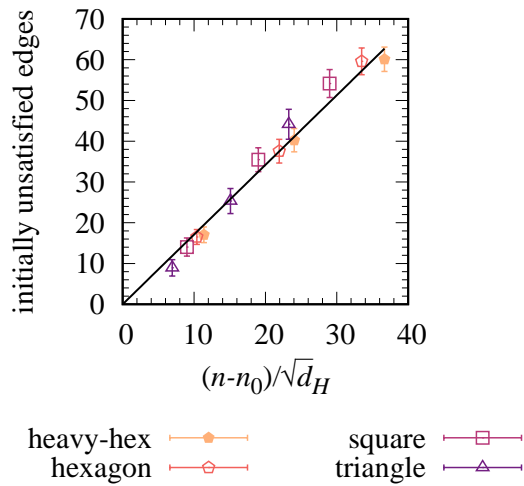}
     \caption{The number of initially unsatisfied edges $N_u$ in the initial qubit placement at each $n$ and $d_H$ for 3-regular graphs. }
     \label{unsatisfied edges}
\end{figure}
\par
For each hardware architecture and circuit, we computed the number of two-qubit edge terms $N_u$ that cannot be implemented directly on the hardware with the initial qubit placement.  The $N_u$ for each hardware are found to scale as $N_u \sim(n-n_0)$, where $n_0$ is a threshold size at which all graphs can be mapped directly to the hardware. The quantity $n_0$ sets the zero of $N_u$ and hence $N_\swap$, for example, on fully connected hardware $n_0 = n$ so $N_u=0$ and no $\swap$ gates are needed.  The rationale for the $n$ dependence is that, on average, the number of unsatisfied edges increases linearly with the total number of edges, $E = 3n/2$ for the 3-regular graphs.  The linear relations $N_u \sim(n-n_0)$ for each individual hardware are shown in Supplemental Information Sec.~I.  They can be related to one another with a factor $d_H^{-1/2}$ that decreases the number of unsatisfied edges when more two-qubit connections $d_H$ are available on the register. This gives a single unified relationship $N_u(n,d_H) \sim (n-n_0)/\sqrt{d_H}$ for all our hardware architectures as shown in Fig.~\ref{unsatisfied edges}.  This motivates and accounts for a factor $(n-n_0)/\sqrt{d_H}$ in the empirical formula in Eq.~(\ref{Nswap 3 regular}).
\par
The remaining factor $\sqrt{n/d_H}$ in the empirical $N_\swap(n,d_H)$ of Eq.~(\ref{Nswap 3 regular}) relates to the average numbers of $\swap$ gates per unsatisfied edge $\overline{N_\swap^{(u)}}$. We can rationalize the $\sqrt{n}$ dependence by considering how many $\swap$ gates are needed to bring qubits together to satisfy an edge $u$, based on the typical distance between qubits on the approximately $\sqrt{n} \times \sqrt{n}$ hardware grids with $\sqrt{n} \in \mathbb{N}$.  We begin by considering uniform random placements of logical qubits along a single dimension of length $\sqrt{n}$.  The probability for the first qubit to be at location $i$ is $P_i = 1/\sqrt{n}$, the probability for the second qubit to be at any other location $j$ is $P_j = 1/(\sqrt{n}-1)$, and the average distance between the qubits is $\sum_{i=1}^{\sqrt{n}}\sum_{j=1}^{\sqrt{n}} P_i P_j |i-j| = (n-1)/[3(\sqrt{n}-1)]$. This scales approximately as $\sqrt{n}$. If qubits are placed uniformly at random in two-dimensions and they move along each dimension separately, for example in the square hardware lattice of Fig.~\ref{hardware}(c), then the total distance is twice the distance in a single dimension and this again scales as $\sqrt{n}$.  In reality the qubit placements are optimized instead of uniformly random, but still the length scales as $\sqrt{n}$ in each dimension and this gives some justification for the appearance of $\sqrt{n}$ in $\overline{N_\swap^{(u)}}$.  Finally, we need to account for a factor $1/\sqrt{d_H}$ to obtain the desired relation  $\overline{N_\swap^{(u)}} \sim \sqrt{n/d_H}$. We rationalize this factor by considering that fewer $\swap$ gates are needed to move a qubit from one location to another when there are more connections $d_H$ on the register, for example, in the triangle lattice some diagonal movements are allowed on the planar grid and we expect this to decrease the number $\swap$ gates that are needed.  We incorporate this through a factor $\sim 1/\sqrt{d_H}$ such that $\overline{N_\swap^{(u)}} \sim \sqrt{n/d_H}$. Combined with the previous analysis of $N_u$, we have $N_\swap(n,d_H) = N_u\overline{N_\swap^{(u)}} \sim (n-n_0)\sqrt{n}/d_H$, giving the empirical formula of Eq.~(\ref{Nswap 3 regular}).  
\acknowledgements
This work was supported by the Defense Advanced Research Project Agency ONISQ program under award W911NF-20-2-0051. J.~Ostrowski acknowledges the Air Force Office of Scientific Research award, AF-FA9550-19-1-0147. G.~Siopsis acknowledges the Army Research Office award W911NF-19-1-0397. J.~Ostrowski and G.~Siopsis acknowledge the National Science Foundation award OMA-1937008.

\section*{Data Availability}
Data from this study is available at https://code.ornl.gov/5ci/dataset-scaling-qaoa-on-near-term-hardware/

\section*{Author contributions}
P.C.L.~contributed to computations, design, analysis, and writing of the manuscript. T.N.~contributed to computations, analysis, and writing. A.S.~and A.M.~contributed to computations. R.H.,~J.O.,~G.S.,~and T.S.H.~contributed to design and analysis; R.H.~ and T.S.H. also contributed to writing.

\section*{Competing interests}
The authors declare no competing interests.

\bibliographystyle{unsrt}
\bibliography{references}

\newpage
\section*{Supplemental Information}
\section{Circuit mapping computations}
We map QAOA problem instances to hardware circuits using the Enfield software library \cite{Siraichi2019Enfield} implemented within the XACC programming framework \cite{McCaskey2018XACC,McCaskey2019XACC}.  We assessed performance of the circuit mapping algorithms WPM, CHW, BMT, and SABRE on example problems, finding that SABRE  \cite{Li2019tackling} gave superior performance and time-to-solution scaling with problem size. We therefore used SABRE for all of our circuit mappings. We optimized two adjustable parameters of the SABRE algorithm to minimize gate counts for each of our test sets.  The first parameter ``iterations" determines how many times SABRE generates random initial placements.  Each of these placements is optimized by SABRE, which outputs the placement with the smallest depth as the final result. The second parameter used by SABRE is called ``lookahead" and determines a balance between current and future gates in the objective function for varying steps in the algorithm, see Ref.~\cite{Li2019tackling} for details.  
\par
\begin{figure}
    \centering
        \includegraphics[width=8.5cm,height=9cm,keepaspectratio,trim={0cm 0cm 0cm 0.cm }]{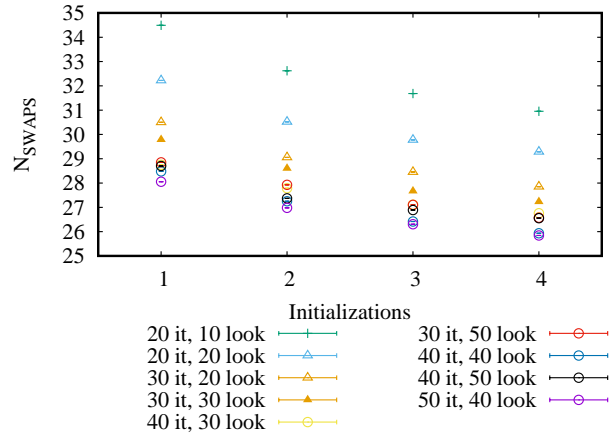}
    \caption{Example convergence behavior of SABRE for variations in the ``iterations" (its) and ``lookahead" (look) parameters, for 3-regular graphs at $n=20$ }
    \label{its lookahead fig}
\end{figure}

Figure \ref{its lookahead fig} shows an example of the convergence behavior for the ``iterations" and ``lookahead" parameters for a series of five ``shuffled" initializations, as described in detail in the next section, for the test-set of 3-regular graphs of size $n=20$.  For each initialization, the results show a clear dependence on the values of ``iterations" and ``lookahead", which become steady when both of these parameters are at least forty.  We set each parameter equal to 40 to compute our final results for the test set.  We observed similar behavior for these parameters at $n=40$, with results that appeared convergent when iterations and lookahead are equal.  For $n=60$ we optimize the parameters assuming they are equal based on our results from $n=20$ and $n=40$.  Table \ref{SABRE parameter table} lists all the final values we use for these parameters for each of our test sets.  
\begin{table}[H]
\center
\begin{tabular}{| c | c | c | c | c |}
    \hline
    $n$ & graph ensemble & shuffles & Sabre iterations & lookahead  \\
    \hline
     7 & non-isomorphic & 50 & 20 & 10 \\
     \hline
     20 & 3-regular & 50 & 40 & 40 \\
     \hline
     40 & 3-regular & 50 & 100 & 100 \\
     \hline
     60 & 3-regular & 50 & 140 & 140 \\
     \hline
\end{tabular}
\caption{SABRE parameters.}
\label{SABRE parameter table}
\end{table}
\subsection{Improvements to optimized circuit layouts}
We improve the SABRE circuit layouts by implementing QAOA-specific cancellations of circuit elements.  $\cnot$ gates can be removed  when a $\swap$ gate ($\swap=\cnot_{ij}\cnot_{ji}\cnot_{ij}$) appears next to a trio of gates for a two-qubit cost term ($\exp(-i\gamma_lJ_{i,j}Z_iZ_j) = \cnot_{ij} \rotation_j(2J_{i,j}\gamma)  \cnot_{ij}$).  This gives adjacent and identical $\cnot_{ij}$ gates which can be removed since $\cnot_{ij} \cnot_{ij} = \mathbb{1}$.  Although each $\swap$ gate is defined by a series of three $\cnot$ gates, the net gate cost from adding a $\swap$ gate with a cancellation is only $\sigma=1$ additional $\cnot$ gate, since two $\cnot$ gates are removed in the cancellation.  To increase the number of these cancellations, we defined a ``shuffling" algorithm that rearranges the commuting edge terms $\cnot_{ij} \rotation_j(2J_{i,j}\gamma)  \cnot_{ij}$ in the circuit that we input to SABRE for optimization.  We found this rearrangement also reduces the total number of $\swap$ gates by finding more efficient series of gates for the hardware. 
\begin{figure}
    \centering
        \includegraphics[width=8.5cm,height=8.5cm,keepaspectratio,trim={0cm 0cm 0cm 0.cm }]{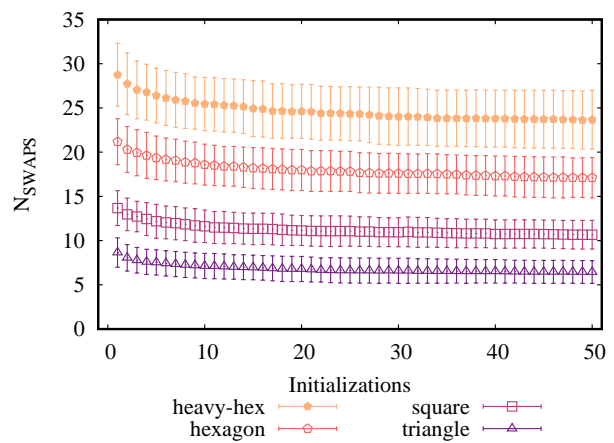}
    \caption{Example convergence behavior with varying shuffled initializations at $n=20$.}
    \label{shuffles convergence}
\end{figure}
\par
We define a shuffling procedure that works in a loop outside SABRE to generate a random ordering for the commuting $\cnot_{ij} \rotation_j(2J_{i,j}\gamma)  \cnot_{ij}$ gate-trios in the circuit.  At each step in the loop, a shuffled QAOA instance is passed to SABRE to determine a final optimized hardware circuit, keeping the circuit with the fewest $\cnot$ gates as the final optimized solution reported in the paper.  Figure \ref{shuffles convergence} shows the convergence behavior of SABRE with additional shuffling iterations. The changes between iterations are very small by about 50 iterations. We find similar behavior for all our test sets, with small changes between subsequent shuffling iterations around 50, hence we use 50 shuffles for all our results.  Each of the 50 shuffle iterations is optimized by SABRE over a number of random qubit initial placements given by the ``iterations" parameter from Table \ref{SABRE parameter table} to identify a final optimized instance.  From the values in the table, this gives 1000-7000 optimizations per graph to identify a single best solution.
\begin{figure}
    \centering
        \includegraphics[width=5.25cm,height=9cm,keepaspectratio,trim={1cm 0cm 2cm 0.25cm }]{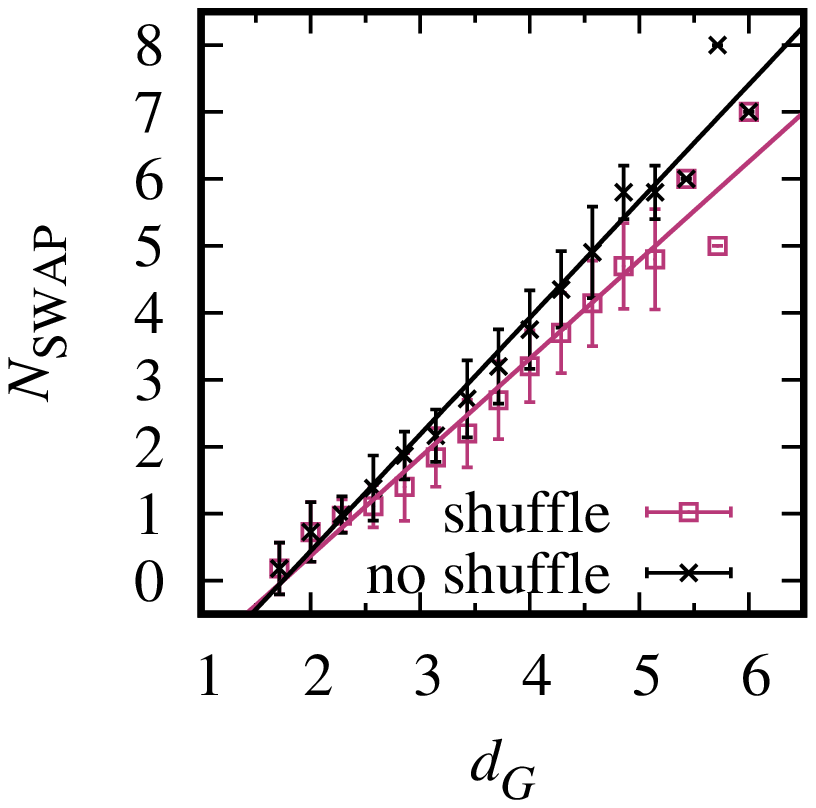}
	\includegraphics[width=5.25cm,height=9cm,keepaspectratio,trim={1cm 0cm 2cm 0.25cm }]{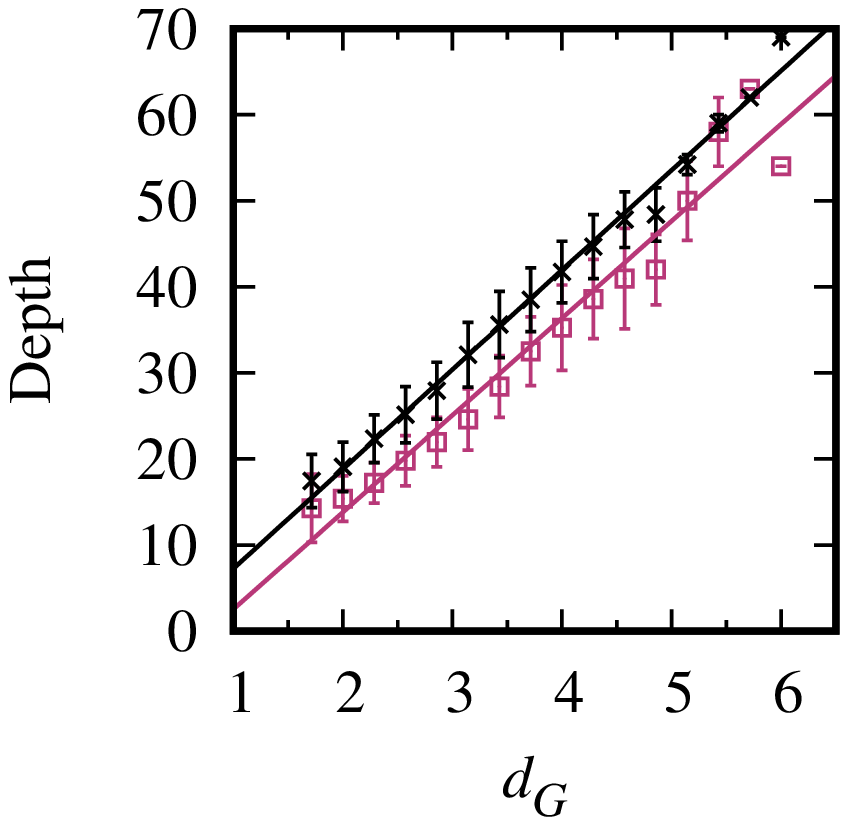}
        \includegraphics[width=5.25cm,height=9cm,keepaspectratio,trim={1cm 0cm 2cm 0.25cm }]{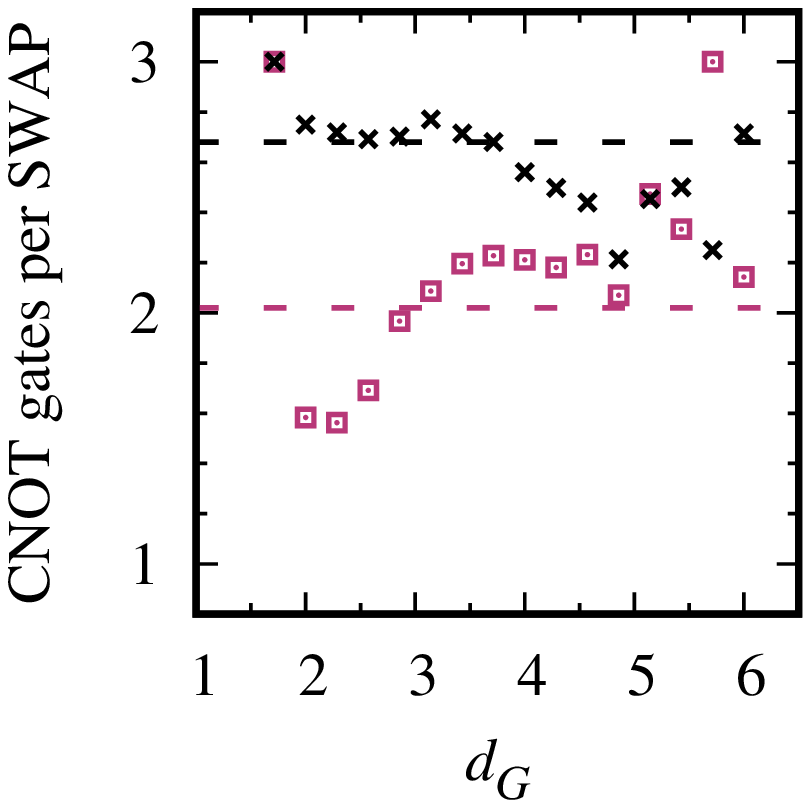}
    \caption{SABRE optimization with and without an exterior loop that shuffles commuting two-qubit circuit components for the edges in the problem graph.  (a) Shuffling decreases the number of $\swap$ gates, (b) the circuit depth, and (c) the net number of $\cnot$ gates per $\swap$ gate $\sigma$ after cancelling pairs $\cnot_{ij}\cnot_{ij}$ in the circuits. }
    \label{shuffling fig}
\end{figure}
Figure \ref{shuffling fig} evaluates the effectiveness of the shuffling iterations relative to an equal number of calls to SABRE without shuffling, for the set of non-isomorphic graphs at $n=7$ mapped to a square hardware grid.  In each figure, the horizontal axis shows the average vertex degree $d_G$ for the graphs, i.e., the average number of non-zero $J_{i,j}Z_i Z_j$ terms per qubit $i$. In the left and central figures, the shuffling algorithm is successful in decreasing the number of $\swap$ gates and the circuit depth, as demonstrated by the linear fits with parameters shown in Table \ref{shuffling table}.  The rightmost figure shows the number of $\cnot$ gates per $\swap$ gate $\sigma$, which is significantly decreased using the shuffling routine, especially at small $d$.  The horizontal dotted lines show the average numbers of $\cnot$ gates per $\swap$ gate, averaged over all graphs with one or more $\swap$ gate.  The average $\sigma$ decreases by about 0.7 when the shuffling routine is implemented.  Overall, the shuffling routine performs well at reducing the QAOA circuit cost by including QAOA-specific circuit commutativity in the SABRE optimization.
     \begin{table}
\begin{tabular}{| c | c | c | c | c |}
    \hline
     & $a$ (shuffle) & $b$ (shuffle) & $a$ (no shuffle) & $b$ (no shuffle) \\
    \hline
    $N_{\swap}$ & $1.47\pm 0.07$ & $-2.6\pm 0.3$ & $1.74 \pm 0.07$ & $-3.0\pm 0.3$ \\
     \hline
     Depth & $11.3 \pm 0.6$ & $-9 \pm 3$ & $ 11.6 \pm 0.3$ & $-4 \pm 1$  \\
     \hline
\end{tabular}
\caption{Fit parameters from the linear fits $f(d) = ad_G+b$ in Fig.~\ref{shuffling fig}. }
\label{shuffling table}
\end{table}
Figure \ref{cnot per swap 3-regular} evaluates $\sigma$ for the sets of 3-regular graphs.  These increase close to $\sigma=3$ as the graph size $n$ increases.  We therefore use  $\sigma=3$ for 3-regular graphs at large $n$ in our scaling analysis. 
\begin{figure}
    \centering
        \includegraphics[width=5.75cm,height=9cm,keepaspectratio,trim={2cm 0cm 2cm 0.25cm }]{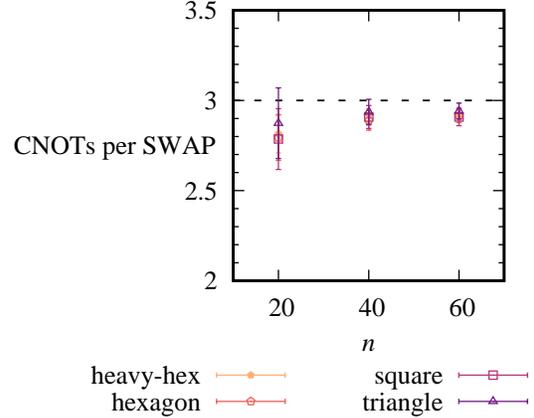}
    \caption{Average increase in $\cnot$ gate counts per $\swap$ gate $\sigma$ in the final circuits for 3-regular graphs.  Values $\sigma<3$ come from cancelling adjacent pairs $\cnot_{ij}\cnot_{ij}$ in the circuits. }
    \label{cnot per swap 3-regular}
\end{figure}
\par
An final improvement comes from cancelling the first layer of $\cnot$ gates in the total QAOA circuit---the initial state is $\hadamard^{\otimes n}\vert 0\rangle^{\otimes n} = \vert +\rangle^{\otimes n}$ and $\cnot_{ij}\vert +\rangle^{\otimes n} = \vert +\rangle^{\otimes n}$, hence the first layer of $\cnot$ gates can be removed.  This gives the factor $N_0$ in Eq.~(5) of the main text.  To systematically search for $\cnot$ gates to cancel in the first layer, we traverse the circuit from left to right and record the operands of each $\cnot$ gate.  If both qubit operands have never been seen before, we cancel the $\cnot$ gate. Note there is an optimal ordering of edge terms ($Z_iZ_j$) to maximize the number of $\cnot$ gates that can be canceled in this way---for example, consecutive disjoint edge terms, such as $Z_0Z_1$ and $Z_2Z_3$, result in more gate cancellation opportunities than a chaining list of terms, such as $Z_0Z_1$ and $Z_1Z_2$.  The procedure allows us to cancel at most $\lfloor n/2\rfloor$ gates, as noted Majumdar \etal \cite{Majumdar2021MaxcutOpt}, since we can have at most $\lfloor n/2\rfloor$ sets of $Z_iZ_j$ gates with disjoint sets of operands $i,j$.  
 \subsection{Degree-based $\swap$ Gate Lower Bound at Small $n$}
 We evaluate SABRE performance in comparison with a simple lower bound for the number of $\swap$ gates for the non-isomorphic graphs at $n=7$. The number of $\swap$ gates can be bounded in terms of the degrees of the hardware connectivity graph and the graph for the cost Hamiltonian, i.e., the set of non-zero $J_{i,j}Z_i Z_j$ terms for the problem instance.  Suppose we have a qubit $j$ with degree $d_G^{(j)} > h_\mathrm{max}$, where $h_\mathrm{max}$ is the maximum degree for a register element on the hardware graph.  In the terminology of the main paper, $h_\mathrm{max}=h$ for the hexagon, square, and triangular hardware lattices, while for the heavy-hexagon it is the maximum number of connections per register element $h_\mathrm{max}=3$.  If $d_G^{(j)} > h_\mathrm{max}$, then at least one $\swap$ will be needed to enable $j$ to interact with additional qubits, since not all of its interactions can be realized directly on the hardware lattice. In one case, the $\swap$ could change places of a qubit $j'$ adjacent to $j$ and another qubit $j''$ that is twice-removed from $j$ such that $j$ and interact with $j''$.  This allows for one additional interaction with $j$.  In a second case, the $\swap$ gate can switch places of $j$ and an adjacent qubit $j'$, which allows up to $h_\mathrm{max}-1$ new connections between $j$ and $j'',j''',$ etc.  So the greatest number of new interactions with $j$ that can be enabled by a $\swap$ gate is $h_\mathrm{max}-1$.  The minimum number of $\swap$ gates that must be performed for a qubit $j$ to allow it to interact with all adjacent vertices in the cost graph is then
\begin{equation} \label{swap min} N_{\swap,j}^\mathrm{min} = \left\lceil \frac{\delta^{(j)}}{h_\mathrm{max} - 1}\right\rceil \end{equation}
where 
\begin{equation} \delta^{(j)} = \left\{ \begin{array}{ccc} 
               d_G^{(j)}-h_\mathrm{max} &:& \hspace{1mm} d_G^{(j)} > h_\mathrm{max}\\
                0 &:& \hspace{1mm} \mathrm{otherwise} \\
                \end{array} \right . \end{equation}
\par
Each $\swap$ gate switches the hardware locations of two logical qubits and thus can enable new interactions for two logical qubits, that is, a $\swap_{ij}$ gate could enable new interactions for both $i$ and $j$.  The minimum total number of $\swap$ gates is then half the sum of $N_{\swap,j}^\mathrm{min}$ for the individual qubits, 
\begin{equation} \label{swap bound} N_\swap^\mathrm{min} = \left\lceil \frac{1}{2}\sum_j N_{\swap,j}^\mathrm{min}  \right\rceil = \left\lceil \frac{1}{2}\sum_j \left\lceil \frac{\delta^{(j)}}{h_\mathrm{max} - 1}\right\rceil  \right\rceil. \end{equation}
\par
\begin{figure}[H]
    \centering
	\includegraphics[width=5.75cm,height=9cm,keepaspectratio,trim={2cm 0cm 2cm 0.25cm }]{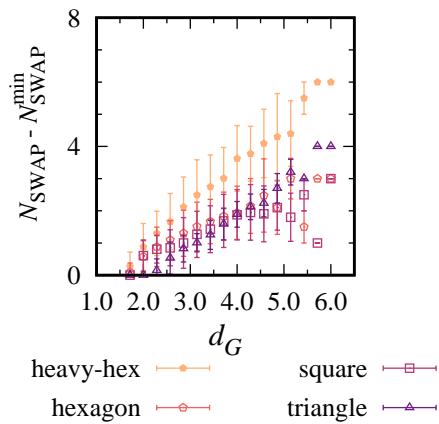}
    \caption{Difference between observed $\swap$ gate counts $N_\swap$ and the lower bound $N_\swap^\mathrm{min}$ for 7-vertex graphs.}
    \label{lower bound comparison}
\end{figure}

Figure \ref{lower bound comparison} compares the observed $\swap$ gate counts for the $n=7$ non-isomorphic graphs to the gate counts from the lower bound.  For each of the fixed-degree hardwares (all except heavy-hexagon), the average observed $\swap$ gate counts are no greater than four above the lower bound.  To be clear, the lower bound is not expected to correspond exactly with the results, since it makes the simplistic assumption that every $\swap$ gate enables the maximum possible number of useful new interactions. Thus we expect the observed gate counts to be higher and take these results as suggesting that SABRE is achieving good performance for these graphs. The agreement is somewhat worse for the heavy-hexagon lattice, which has a mixture of vertices of degree two and three in the hardware graph.  For this we simply use $h_\mathrm{max}=3$ in computing the lower bound, which ignores additional SWAPs required by the degree-two vertices, so worse agreement is expected.
\section{Initially Unsatisfied Edges}
Figure \ref{unsatisfied edges} shows the average number of two-qubit cost terms $\exp(-i\gamma_lJ_{i,j}Z_iZ_j) = \cnot_{ij} \rotation_j(2J_{i,j}\gamma)  \cnot_{ij}$ for initially ``unsatisfied" problem-graph edges that cannot be implemented directly on the hardware in the initial qubit placement, computed from each of the 3-regular graph test sets, see Methods for details.   The number of unsatisfied edges on each hardware scales approximately as $\sim (n-n_0)$ with fit parameters in Table \ref{unsatisfied edge fit table}.  These curves can be united by introducing a scaling factor $1/\sqrt{d_H}$ to each curve, giving the final fit in the Table and Fig.~5 of the main paper.
\begin{figure}[H]
    \centering
        \includegraphics[width=7cm,height=9cm,keepaspectratio,trim={2cm 0cm 1cm 0.25cm }]{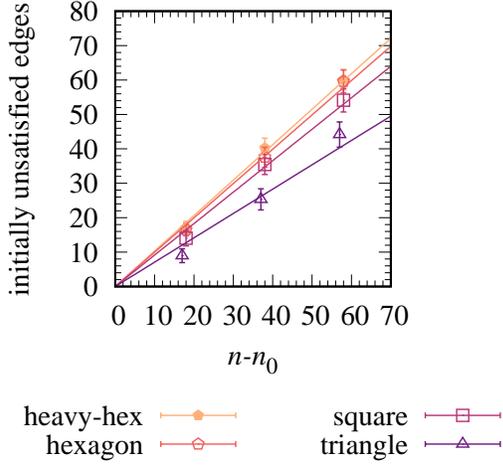}
    \caption{The number of initially unsatisfied edges for the 3-regular graph test sets on each of the hardwares, with fitting functions and parameters of Table \ref{unsatisfied edge fit table}.}
    \label{unsatisfied edges}
\end{figure}
\begin{table}
\begin{tabular}{|c|c|c|c|}
\hline
hardware & $n_0$ & fit function & fit parameter \\
\hline
heavy-hex & 2 & $f_{hh}(n) = \nu_{hh}(n-n_0)$ & $\nu_{hh} = 1.03 \pm 0.02$\\
\hline
hexagon & 2 & $f_h(n) = \nu_{h}(n-n_0)$ &  $\nu_h = 0.99 \pm 0.03$\\
\hline
square & 2 & $f_s(n) = \nu_{s}(n-n_0)$ & $\nu_s = 0.91 \pm 0.04$\\
\hline
triangle & 3 & $f_t(n) = \nu_{t}(n-n_0)$ & $\nu_t = 0.71 \pm 0.06$\\
\hline
all & 2 or 3 & $f(n) = \nu(n-n_0)/\sqrt{h}$ & $\nu = 1.71 \pm 0.04$\\
\hline
\end{tabular}
\caption{Fit functions and parameter values for the numbers of unsatisfied edges on varying hardware lattices.  The final column shows the best fit parameter value and associated asymptotic standard error. }
\label{unsatisfied edge fit table}
\end{table}
\section{Measurement Scaling with error rates, QAOA layers, and problem graph degree}
\begin{figure}[H]
    \centering
        \includegraphics[width=5.75cm,height=9cm,keepaspectratio,trim={1cm 0cm 1cm 0cm }]{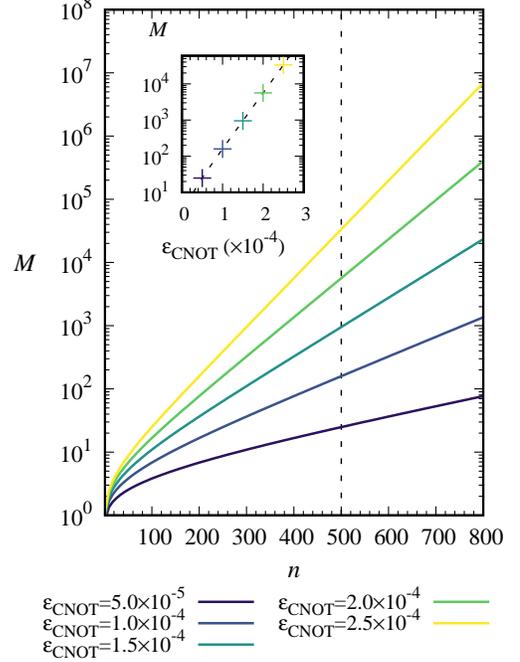}
    \caption{$M$ scaling on fully connected hardware for varying gate infidelities $\epsilon_\mathrm{CNOT}$ with $\epsilon_\hadamard = \epsilon_\rotation = \epsilon_\mathrm{CNOT}/10$. }
    \label{M gate infidelity}
\end{figure}

Figure \ref{M gate infidelity} shows that the number of measurement samples $M$ to obtain a single measurement from the ideal state distribution increases exponentially with the $\cnot$ gate infidelity $\epsilon_\cnot$ on fully connected hardware, with $\epsilon_\hadamard = \epsilon_\rotation = \epsilon_\cnot/10$ as in the main text. Similar scaling can be observed with increasing numbers of QAOA layers $p$, since $F_0 \approx f_0^p$ where $f_0$ is the fidelity lower bound for a single layer.  At small $F_0$ the logarithm $\log(1-F_0) \approx -f_0^p$, so $M \approx -\log(1-\mathcal{P})/f_0^p$ and this diverges exponentially in $p$.  Similarly, increasing $d_G$ increases the number of two-qubit edge terms in the QAOA circuit and the fully connected $N_\cnot^\mathrm{fc} \sim d_G$.  If we further assume the same scaling for $N_\swap$ as for our $n=7$ graphs, then $N_\swap \sim d_G$ and the total number of $\cnot$ gates $N_\cnot \sim d_G$. These factors appear in exponents in $F_0$ and this gives an exponential divergence in $M$ with respect to $d_G$.

\end{document}